# Impact of Distributed Processing on Power Consumption for IoT Based Surveillance Applications

**Barzan A. Yosuf, Mohamed. Musa, Taisir Elgorashi, and J. M. H. Elmirghani**
*School of Electronic and Electrical Engineering, University of Leeds, Leeds, LS2 9JT, United Kingdom*
*E-mail: {elbay@leeds.ac.uk, M.Musa@leeds.ac.uk, T.E.H.Elgorashi@leeds.ac.uk,*
*J.M.H.Elmirghani@leeds.ac.uk}*

**ABSTRACT**
With the rapid proliferation of connected devices in the Internet of Things (IoT), the centralized cloud solution faces several challenges, out of which, there is an overwhelming consensus to put energy efficiency at the top of the research agenda. In this paper, we evaluate the impact of demand splitting over heterogeneous processing resources in an IoT platform, supported by Fog and Cloud infrastructure. We develop a Mixed Integer Linear Programming (MILP) model to study the gains of splitting resource intensive demands among IoT nodes, Fog devices and Cloud servers. A surveillance application is considered, which consists of multiple smart cameras capable of capturing and analyzing real-time video streams. The PON access network aggregates IoT layer demands for processing in the Fog, or the Cloud which is accessed through the IP/WDM network. For typical video analysis workloads, the results show that splitting medium demand sizes among IoT and Fog resources yields a total power consumption saving of up to 32%, even if they can host only 10% of the total workload and this can reach 93% for lower number of demands, compared to the centralized cloud solution. However, the gains in power savings from splitting decreases as the number of splits increases.

## 1. INTRODUCTION

The Internet of Things (IoT) paves the way for a plethora of applications that contribute significantly to enhancing our daily lives, in domains such as security, agriculture and health care, to name a few. It is estimated that, by the end of 2020, the IoT will host between 50 billion and 200 billion devices [1]. One of the classes of IoT devices are IP-enabled smart cameras that are expected to be deployed in future smart cities for real-time surveillance and video analysis purposes [2], [3]. In a news article published in 2013 by The Telegraph, it is reported that there is one surveillance camera for every 11 people in the UK [4]. These devices will generate massive amounts of data and will demand further processing in order to extract knowledge and make informed decisions. In conventional cloud computing, all of the demands must be aggregated and transported over the network to distant data centers housing enormous numbers of servers [5]. This approach can be problematic because it introduces a great pressure on the already over-stressed network, specifically in terms of bandwidth, energy and other resources[1]. There is a growing consensus among academic researchers on the importance of prioritizing energy conservation among other challenges due to its adverse impact on the environment [6]. To address the aforementioned challenges, Fog computing was introduced to extend the cloud functionalities as close as to where the data is being generated, e.g. at the "edge" of the network [7]. The fog computing paradigm proposes for the utilization of idle computational resources available at the "edge" of the network, allowing for the creation of a shared and highly virtualized micro-data center. Having said that, Fog computing is not proposed to replace the Cloud, rather, the goal is to harness the benefits of both paradigms. Here, service requests that are resource intensive, and which may not be able to be entirely run on the constrained Fog devices, is run using the extensive computational power of the cloud data centers. Likewise, those services that can be hosted within the "edge" nodes can avoid the costly overhead of the network and cloud data centers [8]. As depicted in Figure 1, Fog nodes act as intermediary processing servers between the Cloud-DC and the IoT devices. Fog nodes can be deployed anywhere as long as they can have access to connectivity to the internet. Fog nodes can provide storage to cache sensed data, and processing to extract intelligence from it. Devices connected to Fog nodes can be, but not limited to: switches, routers, embedded devices, surveillance cameras, etc [9]. Optical-based networks have been hailed to offer high bandwidths, reduced latency and improved energy efficiency for bit-rate hungry applications, compared to the other technologies such as copper based communication systems or LTE [10]. Our previous work presented energy efficient solutions in cloud and core networks using MILP techniques considering a variety of scenarios including renewable energy, network coding, and virtualization [23] – [37]. The remainder of this paper is organized as follows: In Section 2, the model architecture is described. Section 3 includes performance evaluation and results followed by Section 4 in which we provide our conclusions.

## 2. IOT-BASED SURVEILLANCE MODEL

The architecture shown in Figure 1 consists of four main layers: 1) IoT, 2) Access Fog, 3) Edge Fog and 4) Cloud-DC. The first layer consists of multiple IoT devices that are deployed into different sites, connected to each other and the Cloud through the PON access network. The second and the third layers are part of the PON access network. The ONU devices aggregate the service demands from the IoTs and they are capable of either relaying the aggregated demands onto the next layer or hosting them locally for processing. Likewise, the OLT node acts as the metro node, gateway to the core network as well as being a micro-DC. The fourth layer is the core network linking up multiple core nodes through IP/WDM technologies and providing access to the distant Cloud-DCs. We consider IoT devices such as surveillance cameras that generate requests for processing video streams. These requests are assumed to have the flexibility of being distributed among as many processing nodes as possible. In order for any node to take part in processing a request (partially or fully), the source node must transport the same data to the processing node(s) in question, hence, the bandwidth requirement is part of the request unless the request is hosted locally. In this work, we develop a model using Mixed Integer Linear Programming (MILP). We report the results obtained using the MILP and focus on the impact of distributed processing on the total power consumption. There are two main parts to the total power consumption, which are a) the power consumed due to the network and 2) power consumed due to processing. The sum of these two parts is minimized subject to the flow conservation constraint and the capacity conservation constraints.

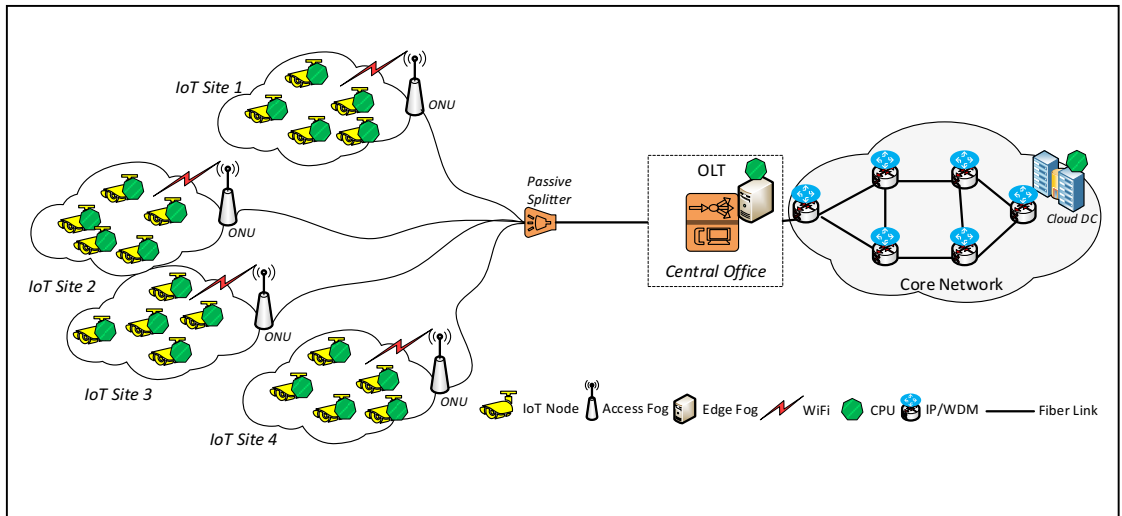

Figure 1: A surveillance application utilizing IoT-based PON networks.

## 3. PERFORMANCE EVALUATION AND RESULTS

Our evaluation scenario considers a typical IoT setting as depicted in Figure 1, where the network consists of 20 IoT devices that are distributed among 4 sites, with each site being connected to a single ONU. A group of the IoT sites is connected to a single OLT device which is located at the network operator's central office (CO) and acts as a metro node [11]. The IoT devices are the sources of the demands and we make the assumption that these demands are flexible in that they can be split among $K$ processing nodes. This assumption is an extension to our previous work in [12]. A service request consists of demands for processing (in MIPS) and bandwidth to transmit data (in Mbps). We assume that all service requests are homogenous among the IoT devices and that the same amount of traffic is transported to all the nodes that process a portion or all of the requested demand. It is assumed IoT devices are smart in nature and thus can process demands locally using their onboard CPUs. In addition, we assume all IoT devices communicate through the WiFi (IEEE 802.11g) protocol with a supported bit rate of 54Mbps. It is important to note that we only consider uplink traffic, as this is reported to carry a much larger proportion of the data generated by IoT networks [13]. Furthermore, we assume the number of CPU instructions required to process one bit of data remains constant throughout our evaluations.

We estimate the CPU intensity (e.g. instructions/bit) per IoT request using data extracted from studies in the literature. In [14], it is reported that 870 instructions/bit are required for object recognition using the traditional Compress-Then-Analyze (CTA) paradigm. Thus, we take a conservative approach and for all of our evaluations we assume 1000 instructions/bit are required to process the video files captured by the IoT devices. As for the bandwidth requirement, we use an online tool to estimate the required data rates for different resolutions and this was estimated to be between 1 Mbps and 10 Mbps, which covers video resolutions between 640×360 to 1280×720 at 10 frames per second [15]. The CPU workload intensity is then calculated by multiplying the instructions/bit by the maximum traffic requirement. Henceforth, this makes the CPU demand proportional to the size of the traffic due to the assumption that, the higher the traffic, the more features a video file will hold, thus more CPU instructions are required to process that file.

We also estimate the processing capacities (in MIPS) of all of the devices, using the data extracted from the technical report in [16]. It is reported that high-end Intel CPUs execute 4 instructions per clock cycle (Hz). We believe it is reasonable to assume that IoT type CPUs execute a single instruction per Hz hence, the number of instructions executed per Hz is multiplied by the device's clock frequency (GHz), in order to estimate the total capacity in MIPS. It is expected that the higher in the network hierarchy of fog-based framework a node is, the larger its computational power due to the assumption that it should provide resources for a larger set of connected devices in the lower layers of the hierarchy [6]. Thus, we assume the Access Fog nodes' processing capacity is more than double that of the IoT's, hence each Access Fog is assumed to execute 2 instructions per Hz. We use the data of the Raspberry Pi Zero W and 2 as representative examples of IoT and Access Fog devices, respectively. Networking devices that are highly shared (such as OLTs/metro nodes, core nodes) [10], are assumed to adopt a fully proportional power profile in which idle power consumption is not considered. This is a reasonable assumption because these devices are already activated by other services such as residential broadband users. However, the rest of devices of interest (networking and processing) that are dedicated to the IoT are assumed to consume idle power. The network within the core layer consists of two layers, the IP layer and the optical layer. In the IP layer, an IP router is deployed at each node to aggregate access network traffic. The optical layer is used to interconnect the IP routers through optical switches and IP/WDM technologies [17]. We assume the power consumption of all of the core network components is fully load proportional and we use the energy per bit approach as described thoroughly in [17]. We also take PUE into account for all of the devices in the network, except those in the IoT and Access Fog layers because these devices do not have any cooling requirements [18]. The results were obtained using AMPL/CPLEX software running on a High performance computer with 16 cores processor and 256 GB of RAM. All of the parameters used in the MILP model, both networking and processing, are summarized in Table 1.

| DEVICE TYPE | | POWER CONSUMPTIONS (WATTS) | | | | CAPACITY | | EFFICIENCIES | | PUE | |
|---|---|---|---|---|---|---|---|---|---|---|---|
| | | Net | | CPU | | Gbps | MIPS | W/Gbps | W/MIPS | Net | CPU |
| | | Max | Idle | Max | Idle | | | | | | |
| IoT (RPi Zero W) | | 0.56[19] | 0.34 | 3.6 [20] | 0.33 [20] | 0.054[21] | 1000 | 4.1 | 0.33 | 1 | 1 |
| Access Fog (RPi 2) | | 15[22] | 9 | 12.5 [20] | 2[20] | 0.3[22] | 2400 | 20 | 0.44 | 1 | 1 |
| Edge Fog | | 48[7] | - | 363[23] | 112[23] | 2.4[7] | 10800 | 20 | 0.023 | 1.5[10] | 2.5[17] |
| IP/WDM node | | 1182[17] | - | - | - | 40[17] | - | 29.6 | - | 1.5[10] | - |
| Metro | Switch | 1766[7] | - | - | - | 256[7] | - | 6.9 | - | 1.5[10] | - |
| | Router | 4550[7] | - | - | - | 560[7] | - | 8.1 | - | 1.5[10] | - |
| Cloud DC | LAN Switch | 1766[7] | - | 363[23] | 112[23] | 256[7] | - | 6.9 | 0.023 | 1.5 [10] | 2.5[17] |
| | LAN Router | 4550[7] | - | | | 560[7] | - | 8.1 | 0.023 | 1.5 [10] | |

*Table 1: Networking and processing parameters used in the MILP model.*

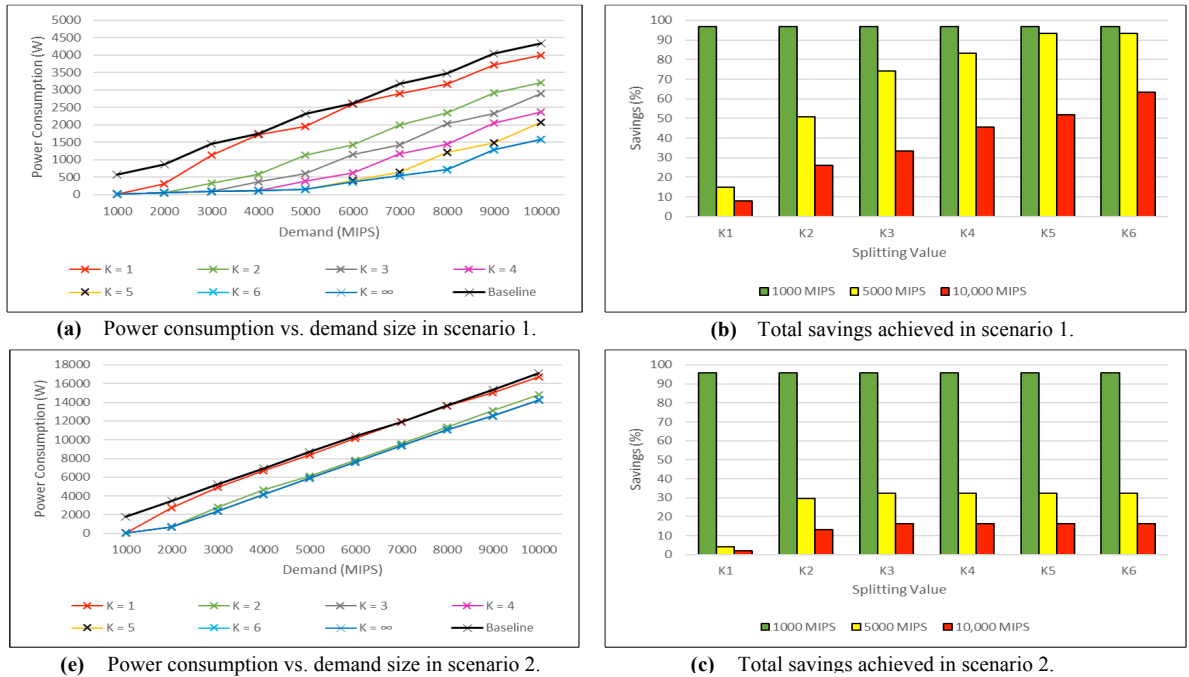

(a) Power consumption vs. demand size in scenario 1.
(b) Total savings achieved in scenario 1.
(e) Power consumption vs. demand size in scenario 2.
(c) Total savings achieved in scenario 2.

*Figure 2: Total power consumptions and power savings under different demand and splitting scenarios.*

Figure 2a and Figure 2b show the total power consumptions and savings in scenario 1 for different workload sizes per IoT device, respectively. In this scenario it is assumed 5 IoT devices are generating demands for services. In Figure 2a, it is shown that, with every increase in the number of splits per IoT demand, savings in total power consumption can reach up to 93%, compared to always processing in the Cloud-DC (baseline). Interestingly, in Figure 2b, it can be observed that for medium sized demands (5000 MIPS) splitting into 2 (*K2*) achieves a total saving of 35% compared to not splitting (*K1*). Whereas splitting the demands into 3 (*K3*) drops the savings down to 24% and at *K4* down to 9%. Also at *K5* and *K6* savings are comparable due to the demands inability to split among the local and Fog layers due to computation capacity constraints.

In the second scenario, it is assumed that all of the IoT devices (20 IoTs) are simultaneously generating requests for computational resources. Likewise, Figure 2c and Figure 2d show the total power consumptions and savings for different workload sizes per IoT device, respectively. It is shown in Figure 2d that splitting medium demand sizes among IoT and Fog resources can still yield a total saving of up to 32%, even if the aforementioned layers can host only 10% of the total workload. It is worthy of mention that in both scenarios, single node placement (K1) achieves negligible savings relative to the savings achieved by splitting onto multiple processing nodes. For instance, in Figure 2d, for high demand sizes (10,000 MIPS), the saving can be as low as 8% and 15% in the first scenario. This is due to the fact that, the lower layers (IoT and Access Fog) cannot accommodate such high demand sizes, thus, the workload must be accommodated by the Edge Fog (Cloudlet) and the Cloud-DC. The networking and processing overheads at these two layers are responsible for the dip in the savings. Also Figure 2d shows that for medium demands, savings are comparable beyond K1 due to the lack of computational resources. However, for higher demands computational resources are exhausted after K2, hence splitting does not achieve any savings.

## 4. CONCLUSIONS

In this paper, we introduced the results obtained using a MILP model which evaluates the impact of distributed processing in an IoT over Fog and Cloud framework. The total power consumption is minimized by partitioning the IoT demands optimally among the available processing nodes. Our results showed that significant savings (93%) in power consumption can be made when the number of IoT requests are low and demands are of medium size. On the other hand, when the number of IoTs increase, a total saving of 32% can still be achieved, even if 10% of the total workload can be located in the IoT and Fog layers.

**ACKNOWLEDGEMENTS**

The authors would like to acknowledge funding from the Engineering and Physical Sciences Research Council (EPSRC), INTERNET (EP/H040536/1), STAR (EP/K016873/1) and TOWS (EP/S016570/1) projects. The first author would like to acknowledge his PhD funding from the Engineering and Physical Sciences Research Council (EPSRC). All data are provided in full in the results section of this paper.